\def\year{2020}\relax
\documentclass[letterpaper]{article} 
\usepackage{aaai20}  
\usepackage{times}  
\usepackage{helvet} 
\usepackage{courier}  
\usepackage[hyphens]{url}  
\usepackage{graphicx} 
\usepackage{graphicx}  
\usepackage{comment}
\usepackage{subcaption}
\usepackage{amsmath}
\usepackage{booktabs}
\usepackage{multirow}
\usepackage{algorithm}
\usepackage[noend]{algpseudocode}
\usepackage{amsmath,amssymb,amsthm,bm}
\usepackage{thmtools,thm-restate}
\DeclareMathOperator*{\argmax}{argmax}

\newtheorem{theorem}{Theorem}

\newtheorem{lemma}[theorem]{Lemma}

\title{Network Inference from a Mixture of Diffusion Models for Fake News Mitigation}

\author{Karishma Sharma, Xinran He, Sungyong Seo, Yan Liu \\ 
Department of Computer Science \\
University of Southern California,
Los Angeles, USA. \\
{krsharma,sungyons,yanliu.cs}@usc.edu}

\begin{document}

\maketitle

\begin{abstract}
The dissemination of fake news intended to deceive people, influence public opinion and manipulate social outcomes, has become a pressing problem on social media. Moreover, information sharing on social media facilitates diffusion of viral information cascades. In this work, we focus on understanding and leveraging diffusion dynamics of false and legitimate contents in order to facilitate network interventions for fake news mitigation. We analyze real-world Twitter datasets comprising fake and true news cascades, to understand differences in diffusion dynamics and user behaviours with regards to fake and true contents. Based on the analysis, we model the diffusion as a mixture of Independent Cascade models (MIC) with parameters $\theta_T, \theta_F$ over the social network graph; and derive unsupervised inference techniques for parameter estimation of the diffusion mixture model from observed, unlabeled cascades. Users influential in the propagation of true and fake contents are identified using the inferred diffusion dynamics. Characteristics of the identified influential users reveal positive correlation between influential users identified for fake news and their relative appearance in fake news cascades. Identified influential users tend to be related to topics of more viral information cascades than less viral ones; and identified fake news influential users have relatively fewer counts of direct followers, compared to the true news influential  users. Intervention analysis on nodes and edges demonstrates capacity of the inferred diffusion dynamics in supporting network interventions for mitigation.

\end{abstract}

\section{Introduction}
\label{sec:introduction}

Falsified information, that is generally intended to deceive people, influence public opinion and manipulate social outcomes has become a prominent topic of discussion. In 2013, the World Economic Forum regarded fake news as a rising global risk in the report entitled `Digital Wildfires in a Hyper-connected World'. Even though deception through falsified information has existed in the past, the increasing use and nature of social media, has made the problem much more intense and difficult to combat.


The risks associated with fake news are more significant due to the scale and reach of social media; the last decade itself has seen more than a ten-fold increase in social media usage \cite{perrin2015social}. The major impacts of fake news have been in social, economic and political issues around the world such as the 2016 US Presidential Elections \cite{allcott2017social}. Besides that, misleading stories discrediting the severity of climate change \cite{roozenbeek2018fake}, and recurrent attempts to promote fear and confusion during natural disasters \cite{gupta2013faking,takayasu2015rumor} cannot be neglected.

Fake news mitigation has been largely studied from the perspective of detection using content analysis, social bots analysis, and analysis of user responses/engagements to the content on social media \cite{sharma2019combating}. As compared to traditional media, online social media allows decentralized dissemination and sharing of content, that can rapidly result in viral information cascades, and widespread impact of misinformation. Therefore, research in intervention strategies to mitigate fake news by monitoring or limiting such diffusions were developed in \cite{farajtabar2017fake,goindanisocial}. \citeauthor{farajtabar2017fake} derived optimal intervention intensities required at nodes in the network to accelerate diffusion of true news through external stimulation. However, facilitating network interventions such as this, requires learning diffusion dynamics of fake and true contents from observed user engagements. Here, we consider the problem of learning diffusion dynamics from observed, but unlabeled cascades of fake and true news; and leveraging the inferred dynamics to facilitate network interventions for fake news mitigation.

\subsection{Contributions and Outline}

In this work, we address the phenomenon of \emph{diffusion} of fake and true contents on social media, using two real-world datasets comprising false and legitimate content cascades collected from user engagements on Twitter. Based on user behaviours in fake and true cascades, we propose a diffusion mixture model (MIC) with parameters $\theta_T, \theta_F$ as a generative model of the diffusion process; and derive \emph{unsupervised} inference techniques for parameter estimation. Unsupervised estimation is important in this domain, since the cost of acquiring labeled (fake/true) cascades is higher due to reliance on expert verification. Using the inferred parameters, we evaluate the role of different users and the network in the propagation of misinformation, and provide analysis for network interventions. The following is an outline of the contributions:

\begin{itemize}
    \item We investigate the nature of user behaviours in response to fake and true news on Twitter datasets. Our findings indicate statistical differences in diffusion patterns of fake and true news with non-homogeneous sharing behaviours.
    \item We propose an unsupervised method to learn the diffusion dynamics from observed, unlabeled information cascades, under the diffusion mixture model MIC, and are the first to examine learnability guarantees in the same.
    \item We evaluate if fake and true cascades are separable using inferred dynamics, compared to unsupervised clustering methods based on text, user and propagation features.
    \item We examine characteristics of users identified as influential in spreading legitimate and fake contents using the inferred diffusion dynamics. Inferred influential fake news users have positive correlation with relative appearance in fake cascades; have relatively fewer counts of direct followers compared to influential true news users; and inferred influential users tend to have engagements in topics of more viral/larger cascades, than smaller ones.
    \item Intervention analysis demonstrates reduction in fake cascade size compared to other unsupervised methods. The learned diffusion dynamics are useful towards actively limiting or mitigating misinformation.
\end{itemize}

\section{Related Work}
\label{sec:relatedwork}

Fake news mitigation is largely addressed as a detection (\emph{classification}) task in existing literature. \citeauthor{sharma2019combating} (\citeyear{sharma2019combating}) classified approaches for fake news detection based on the features used for classification. Broadly, the methods focus on content or writing style analysis \cite{wang2014rumor,khattar2019mvae}, source or bot analysis \cite{ferrara2016rise}, and features from user responses/engagements on social media \cite{qian2018neural,ma2017detect}. The features from user responses are found to be informative and complementary to content or source analysis. In this work, our focus is on information diffusion on social media, to understand how fake and true contents are propagated, and learn a generative model of propagation. \citeauthor{farajtabar2017fake,goindanisocial} (\citeyear{farajtabar2017fake,goindanisocial}) studied intervention strategies based on reinforcement learning for accelerating or limiting diffusions. However, their focus is not on learning diffusion dynamics from observations; and they assume random or known diffusion parameters. Our work on learning diffusion dynamics is therefore complementary to it, and supports different intervention strategies including these.

Network inference refers to the problem of inferring the diffusion process, under a mathematical model of propagation, from observed information cascades. It is studied under different models of propagation \cite{rodriguez2011uncovering,gomez2012inferring,zhou2013learning}. The objective of network inference is to estimate the parameters of a diffusion model from observed information cascades; which might entail inferring the edges of the diffusion network, or both the edges and the strength of influence (or weights) on the edges. For instance, in the Independent Cascade model \cite{kempe2003maximizing}, for every pair of users $u$ and $v$,     there is a parameter $p_{u,v}$ which represents the probability with which $u$ activates $v$, that is information successfully propagates from $u$ to $v$. In other words it is the strength of influence between $u$ and $v$. In the multivariate Hawkes process model, parameters $\alpha_{u,v} \geq 0$ model mutually-exciting nature of network activities, with conditional intensity functions capturing the instantaneous rate of future events conditioned on past events.

Most works in network inference do not address heterogeneity in strength of influence between a pair of users. Furthermore, none of them examine whether the influence is heterogeneous with regards to legitimate and fake contents.  Earlier works only considered topic or time specific networks \cite{yang2013mixture,wang2014mmrate,he2017not}; such as MultiCascades \cite{he2017not} wherein heterogeneous diffusion models are tied together with joint network priors, and inferred from observed but labeled cascades. Our method considers a heterogeneous diffusion model for true and fake news propagation, and in contrast, we propose an unsupervised method for inference that does not require labeled cascades.

\section{Diffusion Mixture Model}
\label{sec:preliminaries}

\begin{figure}[t]
\centering
\begin{subfigure}{.45\columnwidth}
    \centering
\includegraphics[width=\columnwidth]{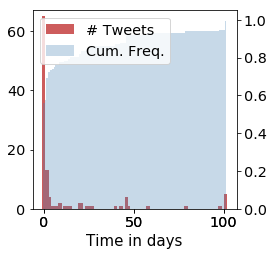}
    \caption{True News Cascade}
    \label{nonlinear_a}
    \end{subfigure}
\begin{subfigure}{.45\columnwidth}
    \centering
    \includegraphics[width=\columnwidth]{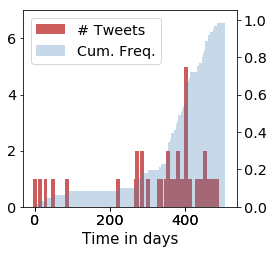}
    \caption{Fake News Cascade}
    \label{nonlinear_b}
    \end{subfigure}
\caption{Example of diffusion cascades on Twitter (\# tweets per day) for (a) emergency landing of an airliner in Hudson river in 2009 (b) information suggesting that the combination of Coke and Mentos can lead to death in 2006.}
\label{fig:intro_cas}
\end{figure}

Information propagation or diffusion is widely studied using probabilistic models, in domains related to viral marketing \cite{domingos2001mining}, and disease and epidemics \cite{newman2002spread}. Diffusion models provide a way to solve important computational problems in each domain. For instance, \citeauthor{domingos2001mining} addressed an important question in viral marketing, that is - to trigger a large cascade of product adoptions, who are the most influential users to target in ad campaigns? Such problems can be efficiently solved using submodular optimization under certain diffusion models such as the Independent Cascade model \cite{kempe2003maximizing}. The choice of model dictates how efficient it is to optimize for important problems such as this. It also affects whether it is possible to derive analytical solutions for learning algorithms in order to infer the parameters of the diffusion model from real observed cascades. Here, we introduce the Independent Cascade Model, followed by our extension of the diffusion model to legitimate and fake cascades.

\noindent \textbf{Cascade:} A cascade is defined as a time-ordered sequence of user responses/ engagements that a piece of information (content) receives, when it is circulated on a social network. It can be labeled as a true or fake news cascade, in accordance with the veracity of the content (eg. Fig.~\ref{fig:intro_cas}).

\noindent \textbf{Independent Cascade (IC) Model:}  First, we discuss the formulation of the Independent Cascade Model studied in \citeauthor{kempe2003maximizing}. $G=(V, E)$ is the directed graph with $n=|V|$ number of nodes (users) and $m=|E|$ edges. A node is \emph{activated} in an information cascade, if its user has an engagement with the content being propagated. Each edge $(u,v) \in E$ is associated with a parameter $p_{u,v} \in [0, 1]$. The diffusion process starts with an initial set of seed nodes assumed to be activated at the first timestep. At each following time step of the diffusion process, a node $u$ activated at the previous time step $t$, independently makes a single activation attempt on each inactive neighbor $v$. The activation succeeds with probability $p_{u,v}$ and a node once activated remains activated in the diffusion process. The influence function $\sigma$ is a function of the seed set $S$ and $\sigma_{\theta}(S)$ is defined as the expected number of nodes activated by the end of the diffusion process starting at seeds $S$, where $\theta = \{p_{u,v} | (u, v) \in E\}$ refers to the parameter set.

\begin{figure}[t]
\centering
\includegraphics[width=\columnwidth=0.9]{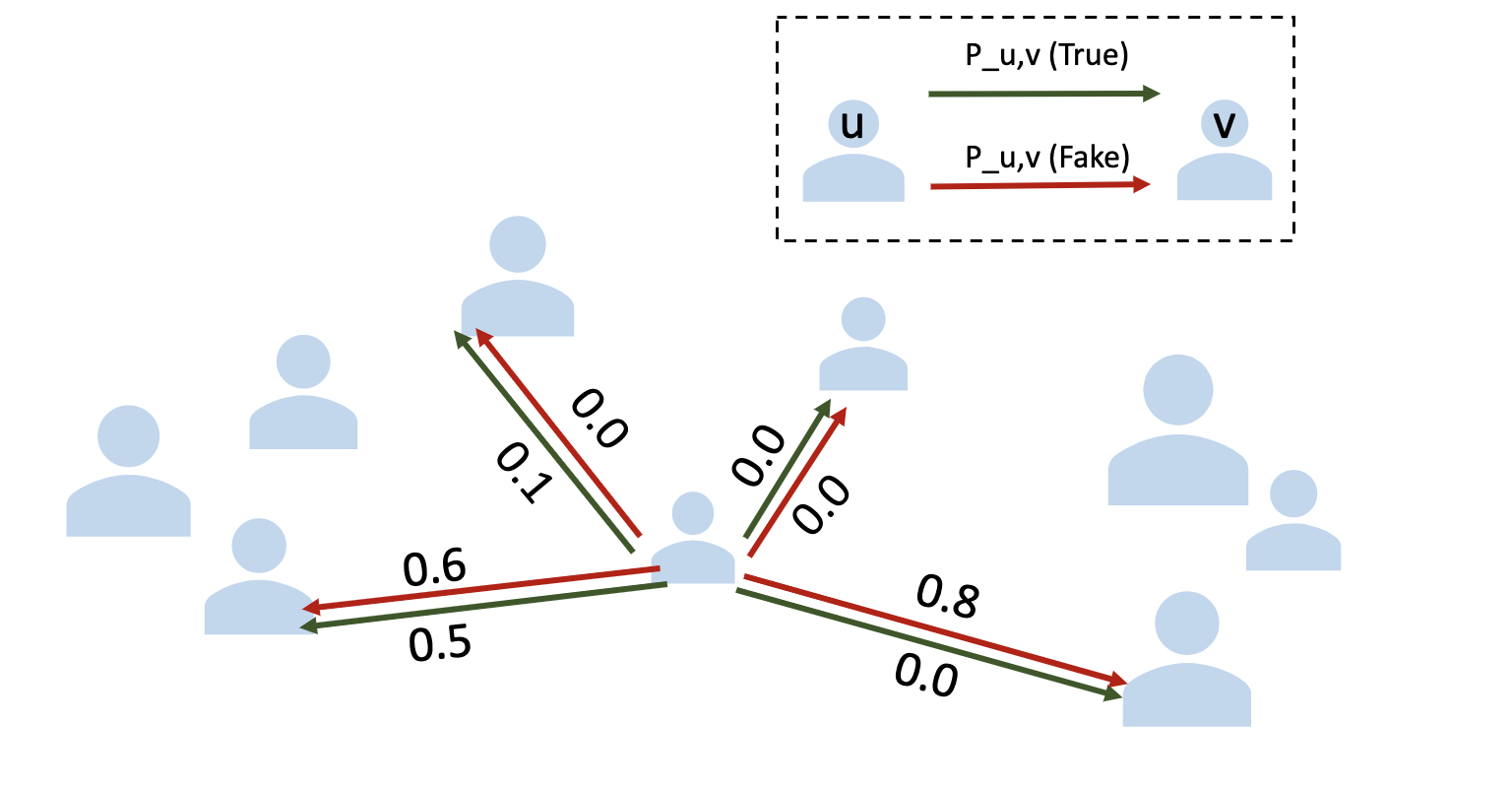}
\caption{Diffusion mixture model (MIC).}
\label{fig:prob_form}
\end{figure}

\subsection{Mixture of Independent Cascade (MIC)}
\label{sec:MIC} 
\label{sec:diffmix} Given a social network $G = (V,E)$, we extend the IC model to include the diffusion of both legitimate (true) and misinformation (fake) contents using separate sets of parameters $\theta_T = \{ p^T_{uv} | (u, v) \in E \}$ and $\theta_F =  \{ p^F_{uv} | (u, v) \in E \}$, i.e., both types of contents share the same network skeleton $G$ but with separate parameters for activation probabilities on the edges. 
Based on this parameterization (illustrated in Fig~\ref{fig:prob_form}), we study the inference of the proposed diffusion mixture model parameters from observed, unlabeled cascades..

First, we formally define the inference problem for the proposed diffusion mixture model, formulated as a mixture of independent cascade models (MIC). We assume that the observed set of diffusion cascades $C$ contains a mixture of unlabeled true and fake cascades. We study whether the diffusion process of true and fake contents can be learned directly from $C$, without requiring cascade labels $\in$ \{true/fake\}. This makes the inference problem more challenging, but practically more useful when collection of labeled cascades requires expert human verification. 

\noindent\textbf{Problem Formulation:} We assume $\pi^T$ is the probability with which a true news cascade emerges, and $\pi^F=1-\pi^T$ is the probability with which a fake news cascade emerges. Let $\pi = [\pi^T, \pi^F]$ be the mixing weights of the diffusion mixture model, then each cascade $c_i \in C$ is assumed to be generated independently under MIC as follows:

\begin{enumerate}
\item Generated seed set $S \subseteq V$ is sampled from some unknown distribution $\mathbb{P}$ over $V$.
\item Generated cascade corresponds to true or fake news based on the outcome of the random variable $h_i \sim$ Bernoulli ($\pi^T$); Cascade labels and mixing weights $\pi$ unobserved.
\item Generated cascade is drawn from the diffusion mixture model with $c_i \sim \textrm{IC}(\theta_T) $ if $h_i = 1$ and $c_i \sim \textrm{IC}(\theta_F)$ otherwise; diffusion parameters $\theta_T, \theta_F$ are unobserved. 
\end{enumerate}
The objective of the network inference problem thereby is to infer $\theta_T$ and $\theta_F$ and $\pi$ from unlabeled cascades $C$.



\section{Real Datasets and Diffusion Analysis}
\label{sec:diffanalysis}

In the previous section, we proposed the diffusion mixture model with separate sets of parameters $\theta_T=\{p^T_{uv}|(u,v) \in E \}$ and $\theta_F=\{p^F_{uv}| (u,v) \in E \}$ for legitimate and fake contents. In this section we first answer two important questions
\begin{itemize}
    \item Are the diffusion patterns of fake cascades significantly different from true cascades?
    \item Are user behaviours with respect to fake and true contents non-homogeneous?
\end{itemize}

We first analyze the diffusion patterns and investigate user behaviours in fake and true cascades. Significant differences between fake and true cascades would mean that the diffusion of fake and true contents are non-homogeneous with respect to user behaviours and should be modeled with separate parameters $\theta_T$ and $\theta_F$. For the purpose of our analysis, we consider real world Twitter datasets described in the following section, followed by statistical hypothesis testing to analyze their diffusion characteristics. A few earlier studies such as \cite{kwon2013prominent,castillo2011information,liu2017rumors} identified which features of a set of hand-crafted features were most discriminative in training classifiers for detecting fake from legitimate contents. There findings indicate that features with high predictive power include - fraction of information flow from low to high-degree nodes which is higher for fake contents, multiple periodic spikes that are particular to fake contents, and greater depth to breadth ratio in the diffusion trees of fake cascades. In our analysis, we consider \emph{temporal} and \emph{structural} differences in diffusion cascades of fake/true news that investigate how user behaviours towards different types of contents differ.




\begin{table}[t]
\centering
\renewcommand*{\arraystretch}{0.7}
\caption{Data statistics for Twitter-1 and Twitter-2.}
\begin{tabular}{@{}l|r|r@{}}
\toprule
Dataset  & Twitter-1 & Twitter-2 \\ \midrule
\# Users                        & 117,824    & 233,719   \\
\# Engagements                   & 192,350   & 529,391   \\
\# Fake Cascades                & 60        & 498       \\
\# True Cascades                & 51        & 494       \\
Avg T length per cascade (hr)   & 8,177   & 1,983     \\
Avg T interval per cascade (hr) & 80     & 65        \\
Avg \# engagements in cascade   & 1,733   & 597       \\ \bottomrule
\end{tabular}
\label{tabstats}
\end{table}

\subsection{Real-World Datasets}
We  utilize  two  publically  available  Twitter  datasets which we refer to as Twitter-1 \cite{kwondata}\footnote{\url{https://dataverse.harvard.edu/dataset.xhtml?persistentId=doi\%3A10.7910\%2FDVN\%2FBFGAVZ}} and Twitter-2 \cite{ma2016detecting} \footnote{\url{https://www.dropbox.com/s/46r50ctrfa0ur1o/rumdect.zip?dl=0}}. Twitter-1 was collected during 2006-2009 and Twitter-2 from March-Dec 2015. In both datasets, topics (contents) are identified as false or legitimate from fact-checking websites like Snopes, and corresponding engagements on Twitter are obtained by keyword search related to the content. The dataset statistics are summarized in Table \ref{tabstats}. For analysis we retain all users, and for inference, we retain users that have at least five engagements in the cascade set, resulting effectively in 3K and 7K users in the two datasets. The former contains $111$ cascades and the later $992$ cascades; with Twitter-1 cascades of average time length of $8177$ hrs and Twitter-2 of $1983$ hrs.

The datasets contain cascades in the form of time-stamped sequences of user engagements, for example, cascade $C_i = [(u_1, t_1), (u_2, t_2), \dots ]$ where $u_j$, $t_j$ corresponds to the engagement of user $u_j$ at time stamp $t_j$ with content corresponding to cascade $C_i$. 
For temporal diffusion analysis, we report statistical tests on each dataset based on the observed cascades. For structural diffusion analysis, we consider only Twitter-1, since it additionally provides follower links from which we can construct retweet structure, similar to \cite{kwon2013prominent}. The follower graph represents whether user A follows user B. Diffusion of a content from B to A can occur if A follows B, and B posts before A in that cascade. Therefore, we can construct the retweet graph of each cascade, from the cascade engagement sequence and the follower graph. In case A has
multiple parents, the edge from the latest parent is retained.

\subsection{Studying User Behaviours in Fake/True Cascades}

In this subsection, we study the diffusion patterns and investigate how these patterns reflect user behaviours. We conduct statistical tests to determine temporal and structural characteristics of fake and true cascades. First, we perform a two-sample $t$-test to verify whether the average time delay between engagements (posts) is higher in fake cascades v/s in true cascades. The first group of samples $S_1$ consists of the fake cascades in the datasets. The second group $S_2$ comprises the true cascades. The log-transform of the data is normally distributed. The null hypothesis is that there is no significant difference between the average time delay between engagements in cascades from the two groups $H_0: \mu_f = \mu_t$. The alternate hypothesis is the average time delay between engagements is higher for fake cascades $H_1: \mu_f > \mu_t$. The $p$-value is shown in Table~\ref{tab:statistical_tests}. The null hypothesis is rejected at significance level $\alpha=0.01$ which suggests that there is statistically significant difference between the temporal characteristics of fake and true cascades.

\begin{table}[t] 
\centering
\renewcommand*{\arraystretch}{0.7}
\caption{Hypothesis testing results ($p$-values) to verify that average time between engagements is higher in fake news cascades than true news cascades (temporal); and to verify that ratio of \# of connected components to total engagements is higher in fake news cascades (structural).}
\begin{tabular}{c cc cc}
\toprule
& \multicolumn{2}{c}{Temporal} & \multicolumn{2}{c}{Structural} \\
\midrule
& $t$-statistic & $p$-value & $z$-score & $p$-value \\
\midrule
Twitter-1 & 4.9975 & $\le$ .00001 & 1.87577 & .03005 \\
Twitter-2 & 12.760 & $\le$ .00001 & NA & NA \\
\bottomrule
\end{tabular}
\label{tab:statistical_tests}
\end{table}

\begin{figure}[t]
\centering
    \begin{subfigure}[t]{0.47\columnwidth}
    \centering
     \includegraphics[height=3cm,width=\columnwidth]{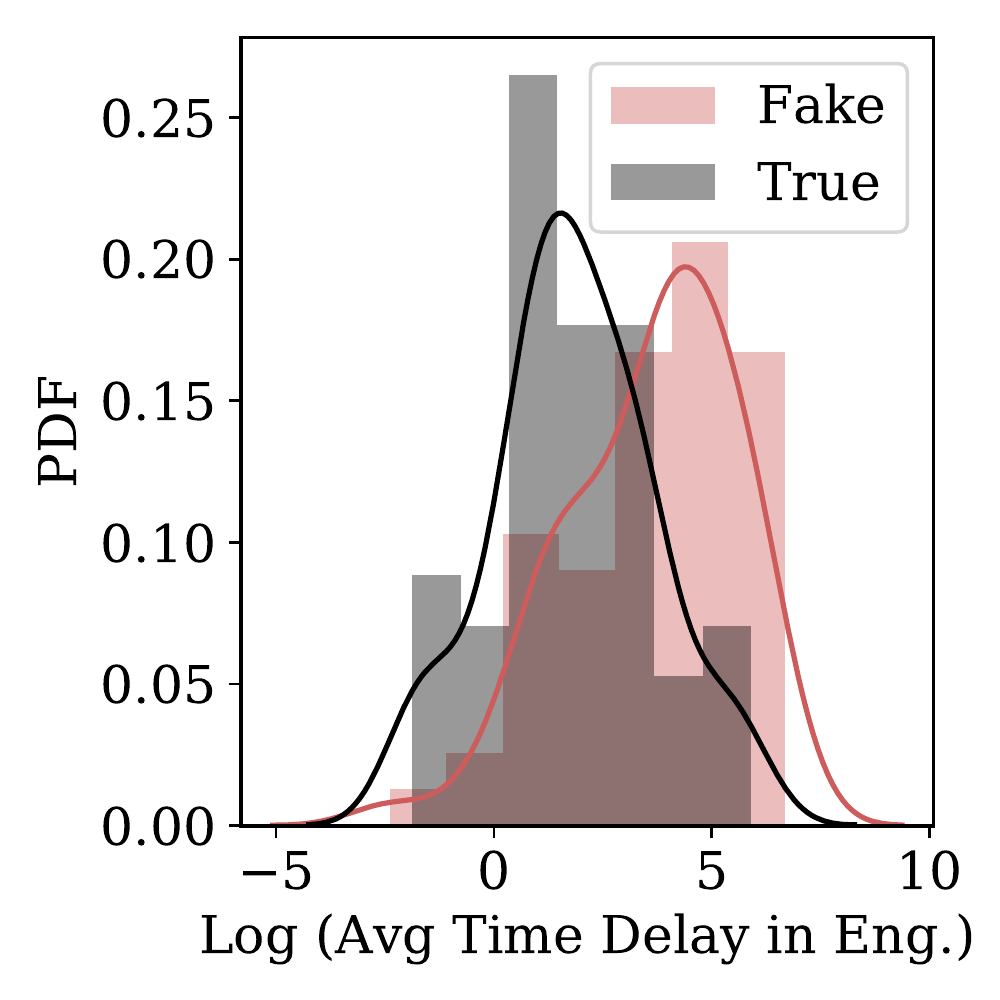}
    \caption{Distribution of avg. time delay after log-transform to reduce skewness for $t$-test in Twitter-1.}
    \end{subfigure}
    ~
    \begin{subfigure}[t]{0.47\columnwidth}
    \centering
     \includegraphics[height=3cm,width=\columnwidth]{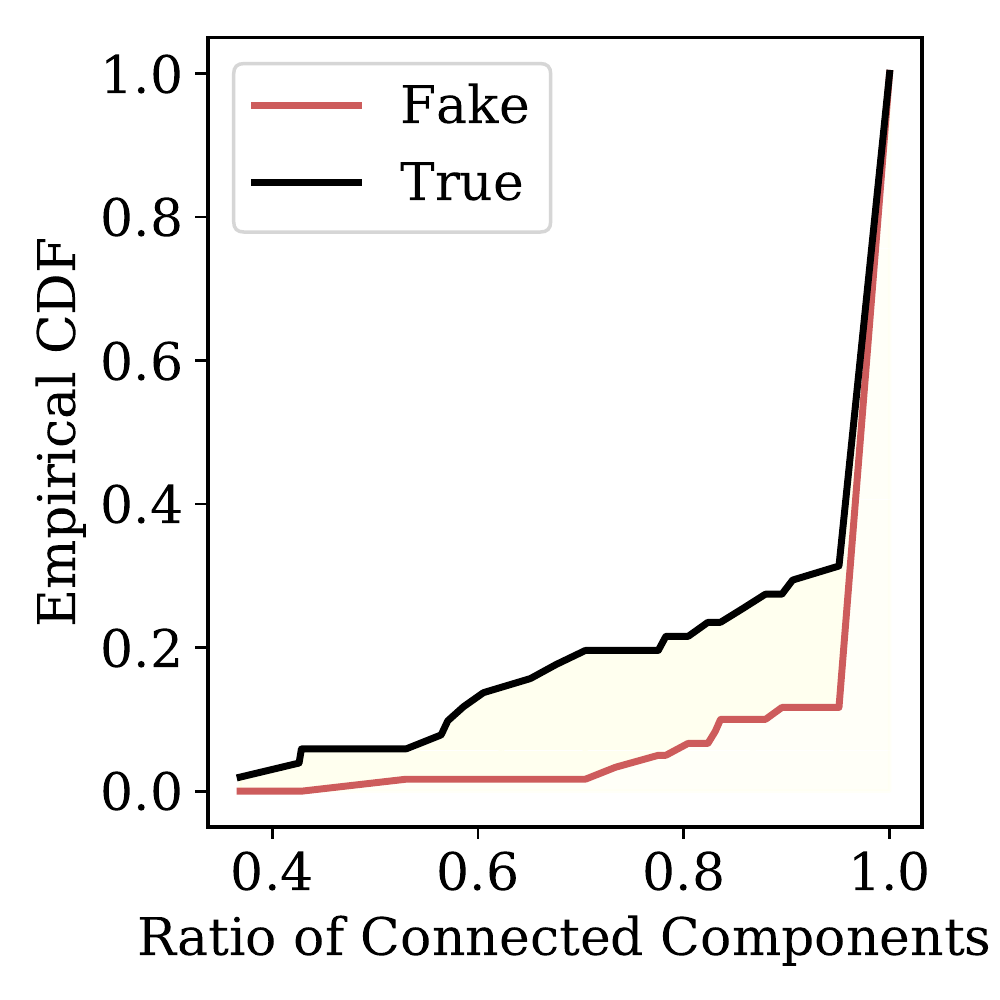} 
    \caption{Empirical CDF of the proportion of connected components in fake/true cascades Twitter-1.}
    \end{subfigure}
    \caption{Statistical tests distributions.}
    \label{fig:stat}
\end{figure}

Second, we perform statistical significance test to examine differences in structural characteristics of the cascades. We compute the number of connected components (cc) in the retweet graph of each cascade, constructed as mentioned in the previous subsection. Then we define the proportion of connected components in a cascade $r = \frac{\textrm{number (cc)}}{\textrm{number of engagements}}$. The null hypothesis is that there is no significant difference in the proportion of connected components in the two groups of fake and true cascades $H_0: r_f = r_t$. The alternate hypothesis is that it is higher for fake cascades $H_1: r_f > r_t$. The data is not normally distributed, so we compute the non-parametric Mann–Whitney U test and report the $z$-score and $p$-value in Table~\ref{tab:statistical_tests}. The null hypothesis is rejected at $\alpha=0.05$ which suggests that the proportion of connected components in fake cascades is higher than in true cascades. Both statistical tests confirm that diffusion patterns differ based on the type of cascade and the user behaviours towards fake and true contents are non-homogeneous. The distribution of average time between engagements and proportion of connected components is provided in Figure~\ref{fig:stat} for Twitter-1. The distribution of avg. time between engagements for Twitter-2 cascades is similar to Twitter-1, and structural follower graph is unavailable in Twitter-2; therefore omitted.

\section{Unsupervised Diffusion Network Inference}

In this section, we consider a reduction of the problem of learning the diffusion mixture under the MIC model to the problem of learning a mixture of product distributions over a discrete domain; followed by an EM algorithm for parameter estimation of the mixture model.

\subsection{PAC-Learnability Reduction}
\label{sec:pac}
Each edge $e=(u,v)$ in the diffusion model is associated with the parameter $p_e^T$ and $p_e^F$ as stated earlier. Each true content cascade can be alternately represented in terms of a `live-edge' graph, such that each edge $e \in E$ is \emph{independently} declared as live with probability $p_e^T$ and included in the graph or blocked with probability $1 - p_e^T$ and not included in the graph. The cascade is then defined by the reachability from seed set S over this graph i.e. a node is activated in the cascade iff there is a directed live edges path from S to the node. Similarly, for fake content cascades. Therefore, each edge can be represented by a random variable $x_e^T$ and $x_e^F$ indicating its live-edge status under the diffusion mixture model, i.e. representing whether the edge $e$ is live or blocked in a given diffusion cascade. Naturally, 
\[x_e^T \sim \textrm{Bernoulli}~~(p_e^T)~~\textrm{and}~~x_e^F \sim \textrm{Bernoulli}~~(p_e^F) \] 

Let $X$ be a vector of random variables indicating live-edge status of each edge under the mixture diffusion model. According to the generative process of the diffusion mixture model, $X$ is a mixture of $k=2$ components $X^T$ and $X^F$ with mixing weights $\pi$. Therefore, $X^T$ is then a discrete distribution over $\{0,1\}^m$  where $m$ is the number of edges in $G$ and all the $x^T_e$ are independent. Therefore $X$ which is the mixture distribution of $X^T$ and $X^F$ is simply a mixture of discrete product distributions with mixing weights $\pi$. The problem is therefore reduced to learning a mixture of discrete product distributions given the live-edge graphs of the observed cascades. Mixture distributions are more generally used in recommendations systems, medicine and other applications \cite{feldman2008learning} and different algorithms can be used to learn the parameters of the mixture distributions, which in our case are $p_e^T, p_e^F$ for all edges in G and mixing weights $\pi$ by definition.

\begin{theorem}{Given a mixture of unlabeled cascades with completely observed live-edge graphs, with diffusion parameters $\theta_M = \{p_e^M|e \in E\}$ with $M \in \{T, F\}$ and any $\epsilon, \delta > 0$, with mixing weight $\pi^M \geq \frac{\epsilon}{mn}$ we can recover in time $\textrm{poly}~$$(m^2n/\epsilon) \cdot \log{(1/ \delta)}$, a list of poly~$(m^2n/\epsilon)$ many candidates, at least one of which satisfies the following bound on the influence function $\sigma_{\theta_M}(S)$ and its estimate $\hat{\sigma}_{\theta_M}(S)$ learned from the observed cascades for seed set $S$ drawn from any distribution $\mathbb{P}$ over nodes in G,
\[
P_{S \sim \mathbb{P}}(|\hat{\sigma}_{\theta_M}(S) - \sigma_{\theta_M}(S)| > \epsilon) \leq \delta \]
with sample complexity $\mathcal{O}\left((\frac{n^4 m^8}{\epsilon^4})^{3} \ln \frac{m}{\delta} \right)$(Proof in Appx).}

\end{theorem}


\subsection{Parameter Estimation}
\label{sec:paramest}

We can estimate the parameters of the diffusion mixture model $\theta_T, \theta_F$ and mixing weights $\pi$ from unlabeled cascades, by deriving a maximum likelihood based estimation procedure. We assume that the observed cascades record the sequence of user engagements, and the order or timestamps of user engagements (activations) are known.

\noindent \textbf{Notation:} \label{sec:notation} We use a general notation $M \in \{T,F\}$ to denote a component in the mixture model, wherein $\{T,F\}$ refer to the true and fake components of the $k=2$ component mixture model MIC. $\theta_T, \theta_F$ are the set of edge influence parameters for each component IC model in the diffusion mixture model with mixing weights $\pi = [\pi^T, \pi^F]$. That is for graph $G=(V,E)$, $\theta_T = \{ p^T_{uv} | u, v \in E \} $ and $\theta_F = \{ p^F_{uv} | u, v \in E \} $. We use the notation $s$ to specify an observed sample cascade belonging to the set of cascades $C$. We define $C_s(t)$ as the set of nodes activated at time step $t$ in cascade $s$ and $t_s(v)$ as the time of activation of node $v$ in cascade $s$. Also, we define $D_s(t)$ as all activated nodes up to and including time $t$. Let $p_{u,v}^M$ and $\hat{p}_{u,v}^M$ as the actual and estimated edge activation parameter in component $M$. In addition, we represent $\gamma_s^M$ as the posterior probability that cascade $s$ is generated under diffusion component $M$ i.e. $\gamma_s^M = P(Z_s = M; \theta)$ where $Z_s$ indicates the component to which the cascade $s$ belongs and $\theta$ is the complete set of parameters $\theta_T, \theta_F$, and $\pi$. Applying Bayes' rule, 
\begin{equation}
\label{eqn:gamma}
\gamma_s^M = P(Z_s = M; \theta) = \frac{\pi^M P(s;~\theta_M)}{\sum_{i=1}^k \pi^i P(s;~\theta_i)} 
\end{equation}
Let $Pa(v)$ represent parents of $v$ in $G$ that is, $u \in Pa(v)$ if and only if $(u, v) \in E$. Similarly, $Ch(v)$ is the children of $v$. Let $p^M_{s}(v)$ be the probability with which v is activated in cascade $s$ under diffusion component $M$. By the definition of the IC model, $v$ is activated at time step $t_s(v)$ in cascade $s$ iff at least one activation attempt of an active parent of $v$ in $s$ is successful. Therefore,
\begin{equation}
\label{eqn:prodeqn}
 p^M_{s}(v) = 1-\prod_{u \in Pa(v) \cap C_s(t_s(v)-1)} (1-p^M_{u,v})
\end{equation}
In addition, let $A_{u,v} \subseteq C$ be the subset of cascades in which both $u$ and $v$ are activated and $t_s(v) = t_s(u) + 1$ and  $B_{u,v}$ be the subset of cascades in which $u$ is activated at some time $t$ and $v$ is not activated up to and including time $t+1$.

\noindent \textbf{Derivation and algorithm:} We derive an expectation maximization based maximum likelihood estimation procedure. The joint log probability of cascade labels and cascades under the mixture model is,
\[
\log P(C, Z; \theta) = \sum_{s \in C} \sum_{M \in  k} 1_{\{Z_s=M\}} \log (\pi^M P(s; \theta_M))
\]
Our goal is to maximize the expected joint log probability,
\begin{align*}
Q &= \mathbb{E} \left[ \log P(C, Z; \theta) \right] \\
&= \sum_{s \in C} \sum_{M \in  k} \gamma_s^M \log \pi^M +  \sum_{s \in C} \sum_{M \in  k} \gamma_s^M \log P(s; \theta_M)
\end{align*}
The maximization of Q with respect to $\pi$ subject to constraints $\sum_{M} \pi^M = 1, \pi^M \ge 0$, we get, $\pi^M = \frac{1}{|C|} \sum_{s}\gamma_{s}^{M}$ from the first term of $Q$ containing $\pi^M$. Now to update the estimates for edge probabilities $p^M_{u,v}$, we need to maximize the second term of $Q$ by differentiating $Q$ with respect to $p_{u,v}^M$. Let $\hat{p}_{u,v}^M$ be the current estimates of edge influence parameters of edge $(u,v)$ for component $M$. As stated in the notations, $C_s(t)$ is the set of activated nodes at times step $t$ in cascade $s$ and $t_s(v)$ is the time of activation of node $v$ in cascade $s$. $p^M_{s}(v)$ is the probability with which $v$ is activated in cascade $s$ under diffusion model $M$. The second term of Q involves the product terms of Equation \ref{eqn:prodeqn} which cannot be solved analytically. However, based on the definition of the IC model, it is possible to approximate Q based on the current estimates of the parameters \cite{gruhl2004information,saito2008prediction}. We utilize the linear approximation chosen in \cite{saito2008prediction}. Primarily Q can be decomposed in terms of nodes activated in a cascade and nodes not activated in a cascade. For the second case of inactive nodes, we will not need any approximation as the likelihood involves $\log (1-p^M_s(v))$ which eliminates the product form of Equation \ref{eqn:prodeqn}. For the first case of active nodes, the form is complex because we do not know which active parent was responsible in activating a given node $v$. This is because, by the definition of IC, activation attempts of all parents of $v$ activated at a given time step are arbitrarily sequenced. Therefore, following \cite{saito2008prediction}, we can instead approximate $p^M_s(v)$ for this case in terms of $\frac{\hat{p}^M_{u,v}}{\hat{p}^M_s(v)}$ for every active parent $u$ since - the probability that $v$ was activated by $u$ should be proportional to the current estimate $\hat{p}^M_{u,v}$ of the strength of influence of $u$ on $v$. Therefore, the second term of $Q$ is as follows, where $X = C_s(t+1) \cap Ch(u)$ and $X' = Ch(u)\setminus D_s(t+1)$,
\begin{equation*}
\begin{split}
\sum_{s \in C} \sum_{M \in  k} \gamma_s^M \sum_{t=0}^{T-1} \sum_{u \in C_s(t)} \Big[ \sum_{v \in X } \Big[ \frac{\hat{p}^M_{u,v}}{\hat{p}^M_s(v)} \log p^M_{u,v} \\
\quad
+ \left(1-\frac{\hat{p}^M_{u,v}}{\hat{p}^M_s(v)}\right) 
\log (1-p^M_{u,v}) \Big] 
+ \sum_{v' \in X'} \log (1-p^M_{u,v'}) \Big]
\end{split}
\end{equation*}
Differentiating the above with respect to $p^M_{u,v}$ and setting it to zero, and considering $Pa(v)$ represents parents of $v$ in base graph $G$, $A_{u,v}$ is the subset of samples in which both $u$ and $v$ are activated and $t_s(v) = t_s(u) + 1$ and  $B_{u,v}$ is the subset of samples in which $u$ is activated at some time $t$ and $v$ is not activated up to and including time $t+1$ we get,
\[
{p}^M_{u,v} = \frac{1}{\sum_{s \in A_{u,v}} \gamma_s^M + \sum_{s \in B_{u,v}} \gamma_s^M} \sum_{s \in A_{u,v}}  \frac{\gamma_s^M  \hat{p}^M_{u,v} }{p^M_s(v)} 
\]
This completes the derivation of the EM procedure with iterative updates in E and M-steps shown in Alg~\ref{algo:em}.


\begin{algorithm}[t]
    \renewcommand{\algorithmicrequire}{\textbf{Input:}}
    \renewcommand{\algorithmicensure}{\textbf{Output:}}
\caption{MIC: Diffusion Mixture Parameter Estimation}

\begin{algorithmic}[1]
\Require observed, unlabeled cascades C
\Ensure estimate $\hat{\theta}_M, \hat{\pi}^M, \gamma_s^M$; \hspace*{0.1cm} $\forall s \in C$ and $M \in \{T, F\}$

\State $\hat{\pi}^T, \hat{\theta}_T, \hat{\theta}_F \gets$ init $\in [0,1]$; $\hat{\pi}^F\gets$ $1-\hat{\pi}_T$.

\While{not converged}
\State // E-Step\;
\State $ \gamma_s^M \gets \frac{\hat{\pi}^M P(s; \hat{\theta}_M)}{\sum_{i=1}^k \hat{\pi}^i P(s; \hat{\theta}_i)}$

\State $p_{s}^M(v) \gets 1 - \prod_{u \in Pa(v)~\cap~C_s(t_s(v)-1)}(1-\hat{p}^M_{u,v})$

\State // M-step\;

\State $\hat{\pi}^M \gets \frac{1}{|C|}\sum_{s \in C} \gamma_s^M$;

\State $\hat{p}^M_{u,v} \gets \frac{1}{\sum_{s \in A_{u,v}} \gamma_s^M + \sum_{s \in B_{u,v}} \gamma_s^M} \sum_{s \in A_{u,v}}  \frac{\gamma_s^M  \hat{p}^M_{u,v} }{p^M_s(v)}$

\EndWhile
\end{algorithmic}
\label{algo:em}
\end{algorithm}

\noindent \textbf{Relaxation:} Since observed cascades only contain the order of activations or time stamps at which users are activated, rather than discrete timesteps, and the edges in $G$ are unobserved; we relax Equation \ref{eqn:prodeqn} to deal with continuous time and let $p_s^M (v)$, the probability that $v$ is active in $s$ under component $M$ equal $1- \prod_{u \in C_s(t_s(v) - W \leq \tau < t_s(v)))} (1-p^M_{u,v})$ where $W$ is a lookback window and hyperparameter of the algorithm. Thus, any $u$ activated in $W$ before $t_v(s)$ is considered a potential parent and influencer (that can activate) $v$. $W$ can be set in unit of time or in terms of number of past events.


\section{Experimental Analysis on Real Datasets} 
\label{sec:experiments}

Using the parameter estimation algorithm, we infer diffusion mixture MIC parameters for the real Twitter datasets described earlier. From the inferred parameters, we evaluate if fake and true cascades are separable based on inferred diffusion dynamics, compared to unsupervised baseline methods for clustering cascades. Next, we identify users that are influential in the propagation of true and fake contents, from the inferred parameters and learned diffusion model; and investigate their characteristic features from the data. Lastly, we demonstrate node and edge interventions based on the inferred diffusion dynamics and show reduction in fake cascade size compared to other baselines.

\subsection{Clustering Cascades}  
From the inferred parameters, we can determine if an observed cascade is more likely to be considered fake or true based on the posterior probability of the cascade under each component of the mixture MIC.
The predicted component for each cascade is thus obtained as $\argmax_M \gamma_s^M$. This will result in two clusters of cascades. Each cluster is assigned fake or true label based on a held out one-fifth set of cascades with known labels. In Table~\ref{tab:cls1},
we evaluate if the fake and true cascades in the datasets are separable based on inferred diffusion dynamics, compared to unsupervised baseline methods for clustering cascades. The implemented baselines are as follows - \emph{TruthFinder} (\textbf{TF}) \cite{yin2008truth} is a credibility propagation algorithm that exploits conflicting sentiments between user comments to the same content. \emph{StanceEval} (\textbf{SE}) exploits the average sentiment of users tweet texts in a cascade as a measure of its type; as it is found that fake cascades tend to elicit negative and questioning responses \cite{qian2018neural,zhao2015enquiring}. \emph{K-Means} \textbf{(KM)} clustering based on temporal and propagation features identified in \cite{ma2015detect,kwon2013prominent,castillo2011information} namely, number of posts in a cascade, time length of cascade, average time gap between posts in the cascade, and fraction of most active users in the cascade. \emph{SEIZ} (\textbf{SZ}) \cite{jin2013epidemiological} is a rumor model proposed for unsupervised rumor detection. It partitions users as either ``susceptible", ``infected", ``exposed" or ``skeptic" with regards to the content and models state transitions between them. The model is fit to each cascade separately by solving differential equations. They define a ratio based on the learned parameters of the rumor model for each cascade to classify it as fake or true. Lastly, we include \textbf{HIC}, where we assume a homogeneous IC model, with a single parameter value (f) shared over all edges for the fake component and another single parameter value (t) for the true component i.e. $p^T_{uv} = t$  $\forall (u,v)$.

 \noindent \textbf{Summary:}
 In comparison with MIC where differences in inferred dynamics are exploited for separation; TF and SE utilize aggregate sentiments of user responses which are relatively noisy signals of veracity. We find that KM was biased towards producing a single cluster without being able to effectively separate them; HIC does not model heterogeneous influence across user pairs which limits expressibility of the model; and SZ cannot capture common patterns across cascades, as it fits separate parameters per cascade. 

\noindent \textbf{Analysis:} In terms of the distribution $\pi$ estimated in MIC, we report the Mean Absolute Error (MAE) between the estimated value and the true data distribution. The data is near balanced, and the estimated $\pi=[0.44, 0.56]$ in Twitter-1 is close to the true distribution, with Mean Absolute Error (MAE) of $0.04$. Twitter-2 estimated $\pi=[0.47, 0.53]$ with MAE of $0.058$. Therefore, MIC outputs balanced clusters of cascade types. KMeans (KM) on the other hand produces unequal sized, biased clusters, resulting in close to random accuracy predicting most cascades to one type, with low f1. The sentiment analysis methods like StanceEval (SE) make mistakes in cases where true content evokes negative sentiments such as ``Is horrified to read about the missing Air France plane'' and also due to sentiment lexicons that map certain words like ``missing" to negative, such as in ``Air France jet missing with 228 people over Atlantic after running into thunderstorms". This results in negatively correlated predictions below 50\% depending on the sentiment patterns and content in the data. TruthFinder (TF) also utilizes user sentiments but is more robust as it accounts for conflicting relationships between users. SEIZ (SZ) is the better baseline based on rumor modeling. But it does depend on an estimated threshold for the ratio per cascade used to determine if the cascade is fake. \cite{jin2013epidemiological} use median ratio over the set of cascades as the threshold, and any observed ratio above this threshold is considered as fake. This can be result in lower quality estimates of the threshold in datasets like Twitter-1 with fewer cascades.

We additionally compare the IC model with MIC. IC model does not have the proposed parameterization for different cascade types, and hence cannot be compared in clustering. Therefore, we report the Average Negative Log-likelihood (NLL) or loss per cascade instead, after parameter estimation using IC and MIC in the datasets. Lower NLL indicates better fit to the observed cascades. Average NLL per cascade on a $20\%$ held-out set of cascades in Twitter-1 for MIC is $9.035$ (train cascades $6.29$), and for IC, it is $3757.84$ (train cascades $58.45$). Average NLL per cascade in Twitter-2, for MIC is $8.45$ (train cascades $8.35$), for IC, it is $1414.25$ (train cascades $1052.97$). Therefore, separate parameters to represent cascade types allows better diffusion modeling.

\begin{table}[t]
\centering
\caption{Clustering (Cascade Separability) Results.}
\renewcommand*{\arraystretch}{0.7}
\begin{tabular}{p{0.5cm}p{1.4cm}p{1.4cm}p{1.4cm}p{1.1cm}}
\toprule
& \multicolumn{2}{l}{Twitter-1} & \multicolumn{2}{l}{Twitter-2} \\ \midrule
 & F1-Score  & Accuracy & F1-Score  & Accuracy \\
\midrule
TF &0.576 & 0.522 & 0.573 & 0.536 \\
SE &0.535 & 0.531  &0.388 &0.469  \\
KM&0.253 &0.522 &0.312 &0.490 \\
SZ &0.54$\pm$0.03 & 0.52$\pm$0.03 &0.56$\pm$0.03 & 0.57$\pm$0.01 \\
HIC&0.48$\pm$0.16 & 0.55$\pm$0.01  &0.49$\pm$0.12 & 0.53$\pm$0.02 \\
\midrule
\textbf{MIC}&\textbf{0.67}$\pm$0.02 & \textbf{0.61}$\pm$0.02 & \textbf{0.63}$\pm$0.01 & \textbf{0.59}$\pm$0.01  \\ \bottomrule
\end{tabular}
\label{tab:cls1}
\end{table}

\subsection{Influential Users Test}
\label{sec:influsers} 


\begin{figure}[t]
\centering
    \begin{subfigure}[t]{0.48\columnwidth}
    \centering
    \includegraphics[height=3.5cm,width=\columnwidth]{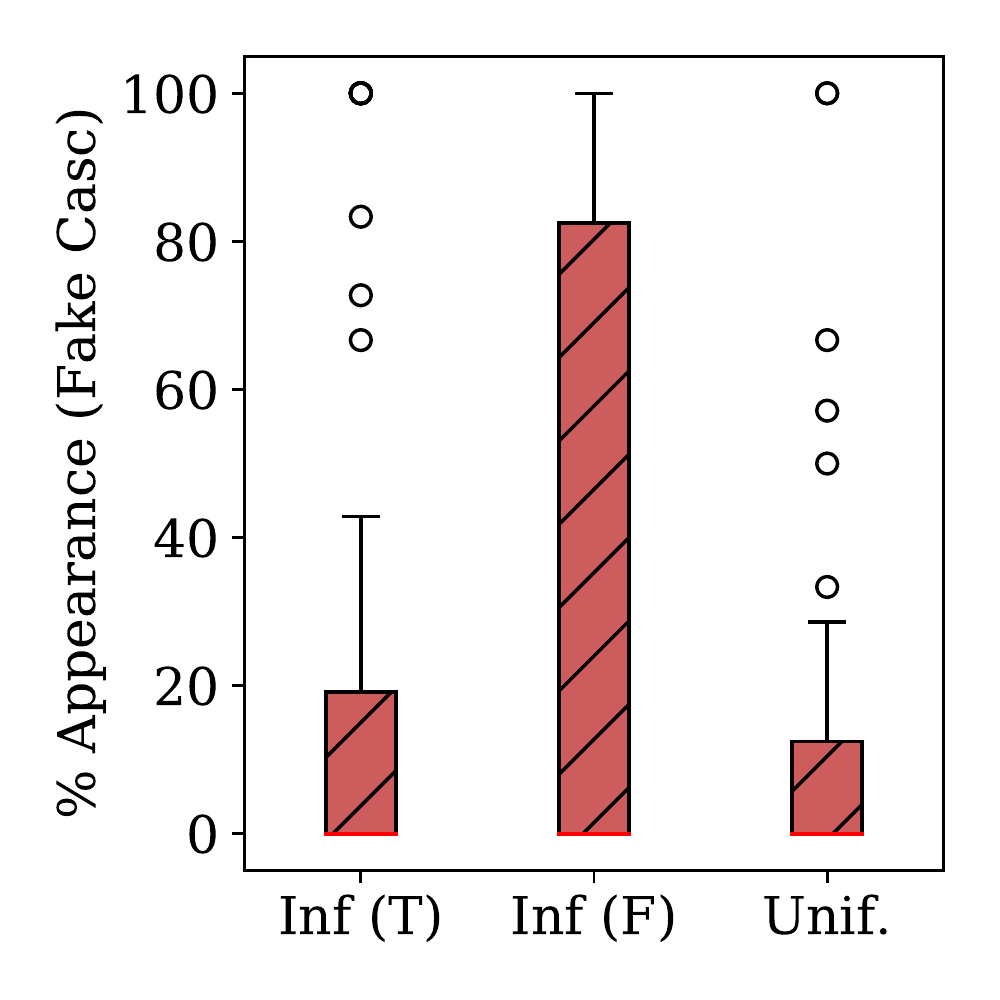}
    \caption{Twitter-1}
    \label{fig:user-analysis-kwon}
    \end{subfigure}
~
    \begin{subfigure}[t]{0.48\columnwidth}
    \centering
    \includegraphics[height=3.5cm,width=\columnwidth]{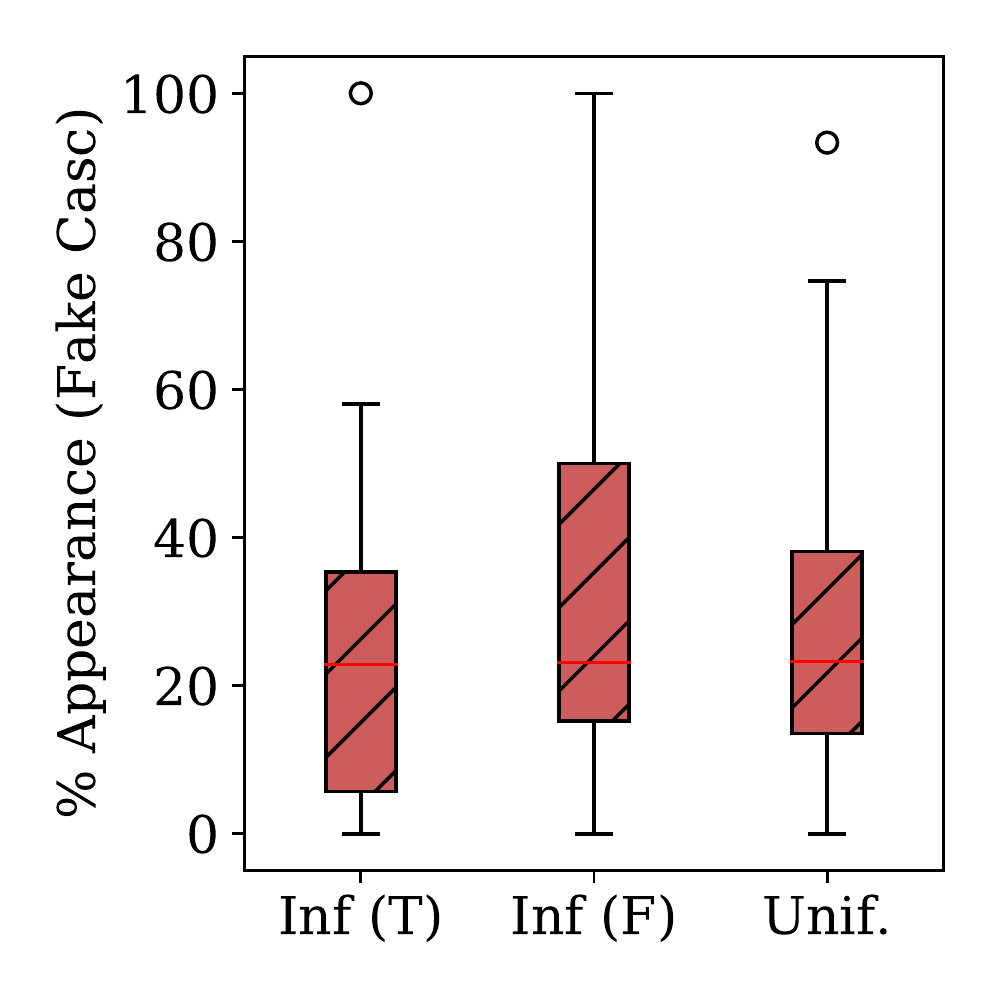}
    \caption{Twitter-2}
    \label{fig:user-analysis-ma}
    \end{subfigure}
\caption{Results on quality of influential users selected based on the estimated diffusion parameters for true and fake news in (a) Twitter-1 and (b) Twitter-2. Inf(T) and Inf(F) are inferred influential users for true and fake news.}
\label{fig:user-analysis}
\end{figure}

In this subsection, we first identify users that are influential in the propagation of true and fake contents (i.e. users that if selected as seed sets would trigger the largest cascades), using the inferred diffusion model. Selecting and analyzing the top $100$ influential  users identified for fake and true news, we can further evaluate the quality of inferred parameters.

\noindent \textbf{Selection of influential users:} Influential users for each component IC model with inferred params $\hat{\theta}_T, \hat{\theta}_F$, are selected using greedy maximization algorithm implemented based on \citeauthor{goyal2011celf++} \citeyear{goyal2011celf++}.

\noindent \textbf{Result:} In Fig~\ref{fig:user-analysis}, we report the box-plot for inferred influential users based on \% relative appearance in fake vs true cascades. Inferred users identified for fake news (Inf(F)) have high positive correlation with relative appearance in fake news cascades, as seen from the figure, for both datasets, in comparison with influential true news users (Inf(T)), and a uniform random sample of users (Unif.).

\noindent \textbf{Degree of separation:} Uniform random sample (Unif.) of users, provides insights into the degree of separation between true and fake cascade clusters in the two datasets. Compared to Twitter-1, uniformly sampled users in Twitter-2 are more likely to engage with both contents, whereas in Twitter-1 the median of the uniformly sampled users interact purely with true contents; showing potentially larger separation in Twitter-1 between fake and true cascade users.

\begin{table*}[t]
\centering
\caption{Characteristics of most influential users inferred for propagation of fake and true news in Twitter-1.}
\renewcommand*{\arraystretch}{1.0}
\begin{tabular}{llllll}
\toprule
Comp & \# Followers & \# Following & \# Posts & &  Description Tags
\\
\midrule
\multirow{5}{*}{True} &
54418 & 1157 & 24182 & HuffPost & real life is news, and news is personal. Read more: https:..

\\

& 17874	& 0	& 675 & TMZ & breaking biggest stories in entertainment news 

\\
& 2684 &	2941	& 2144 & $\_$PSPGuru & Sending you constant news about the latest PSP news.
\\
News &
1118 & 142 & 3191 & FOX10News
& TV news station, serving the Alabama, Florida, Gulf 
\\
& 22252 &	23853 &	4621 & OnlyMobileNews & We follow the latest in mobile technology news

\\
\midrule
\multirow{5}{*}{Fake}  & 
672	& 280 & 8200 & unk. & \textbf{F.} SwineFlu from pork, SwineFlu zombies \textbf{F.} 45
\\

 & 
514	& 470	& 3408 & Terrypooch & Fighting for liberty and justice for all \textbf{F.} 308
\\

& 
1 & 0 & 32 & 08kx250f  &  \textbf{F.}  xbox720 will launch before 2012 \textbf{F.} 9
\\

News & 3273 & 1926 & 2294 & unk. & \textbf{F.} Obama is not a natural born citizen \textbf{F.} 5
\\


& 8829 & 8362 & 6375 & BuzzFeed  &  \textbf{F.} BigFoot, GiantCatfish Maneater, MontaukMonster \textbf{F.} 7

\\
\bottomrule
\end{tabular}
\label{tab:userfeats}
\end{table*}

\begin{figure*}[t]
\centering
    \begin{subfigure}[t]{0.47\columnwidth}
    \centering
    \includegraphics[width=\columnwidth]{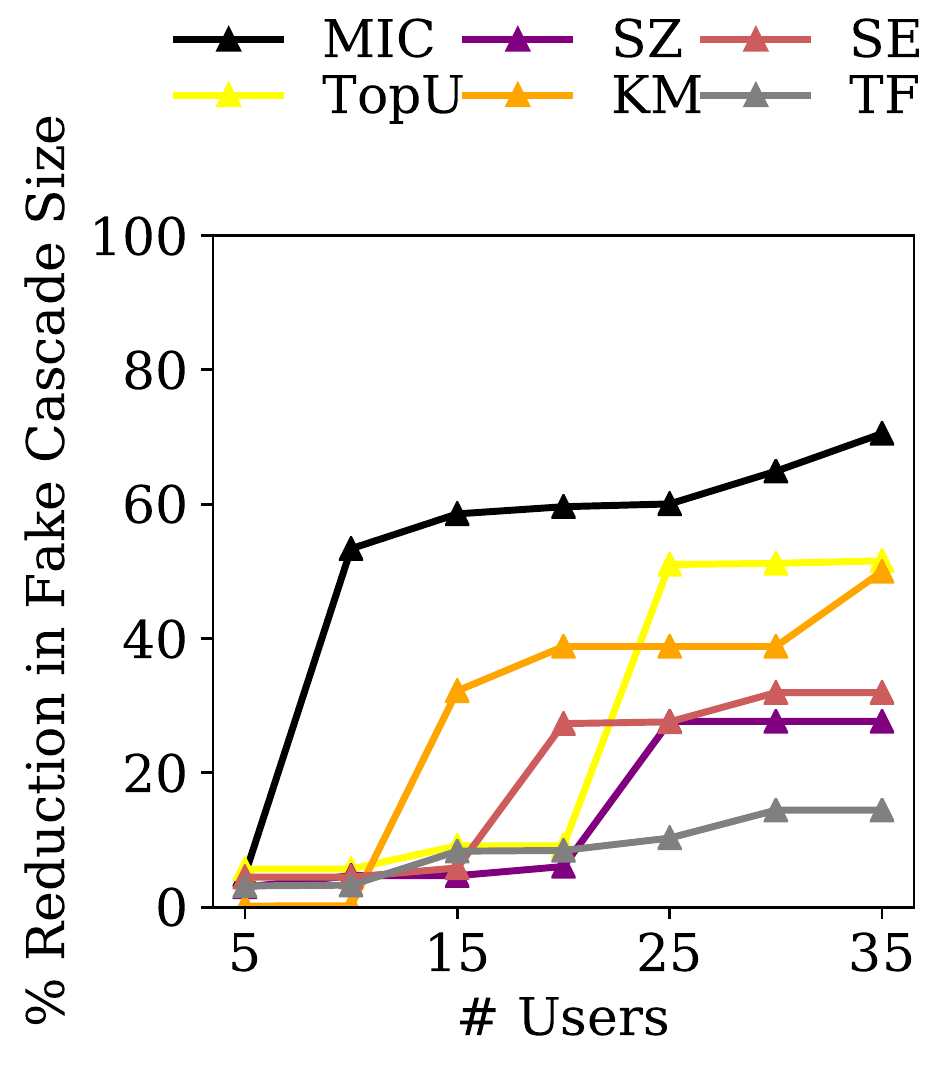}
    \caption{Intervention: Twitter-1.}
    \label{fig:ninter_kwon}
    \end{subfigure}
    ~
    \begin{subfigure}[t]{0.47\columnwidth}
    \centering
     \includegraphics[width=\columnwidth]{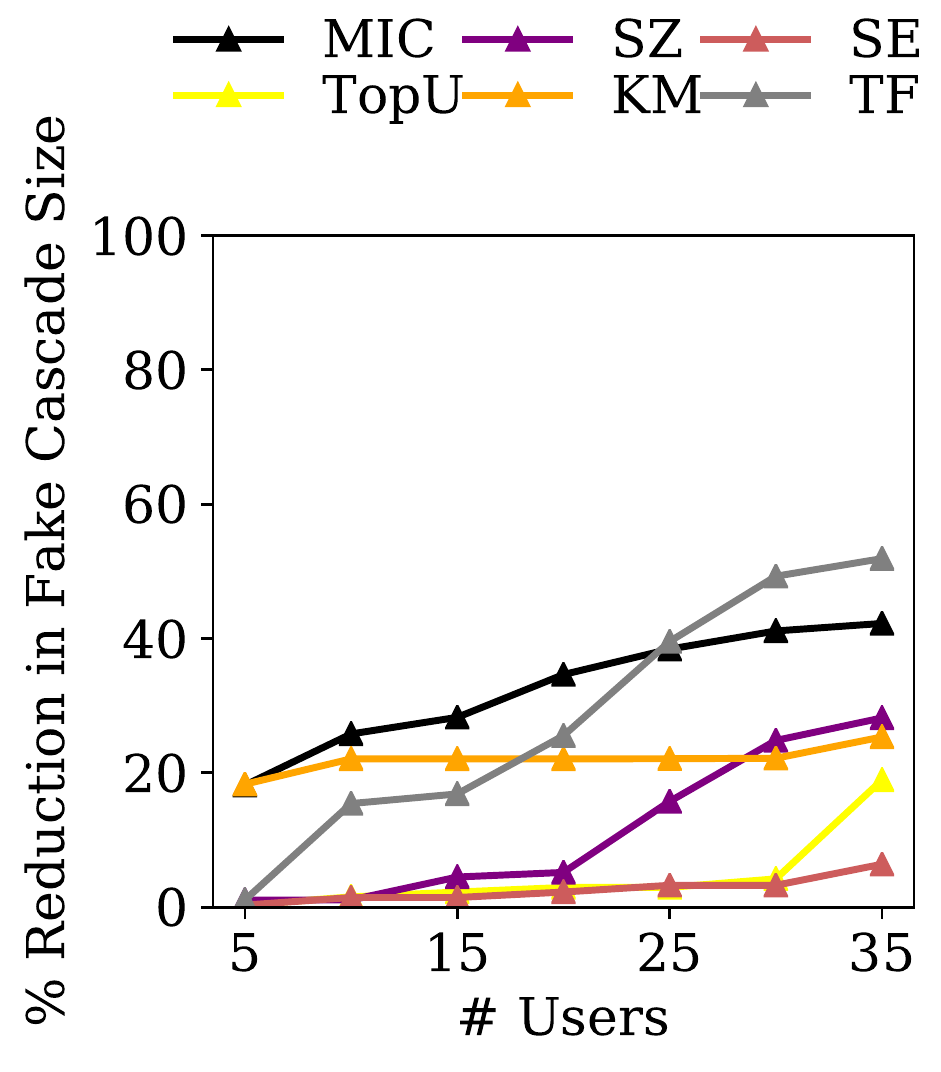}
    \caption{Intervention: Twitter-2.}
    \label{fig:ninter_tma}
    \end{subfigure}
    ~ \hfill
    \begin{subfigure}[t]{0.5\columnwidth}
    \includegraphics[width=\columnwidth]{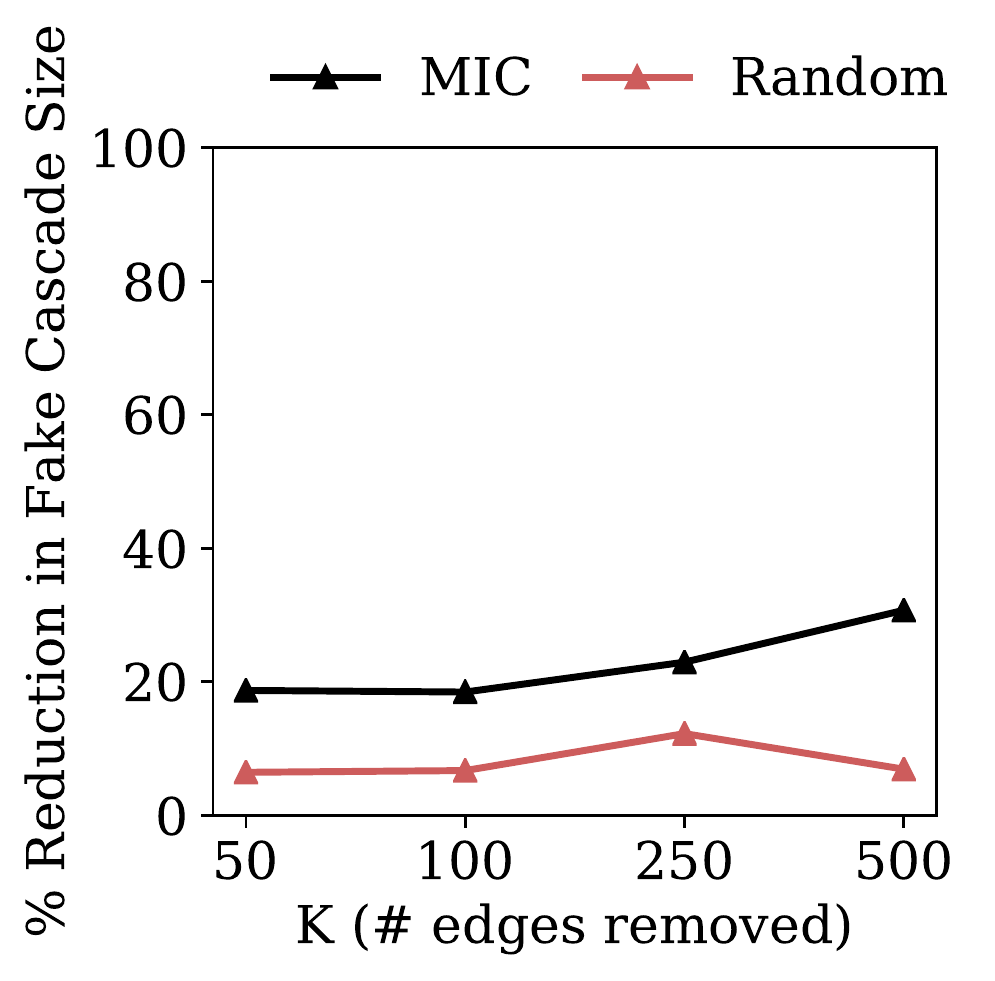}
    \caption{Intervention: Twitter-1}
    \label{fig:einter_kwon}
    \end{subfigure}
    ~
    \begin{subfigure}[t]{0.5\columnwidth}
    \centering
     \includegraphics[width=\columnwidth]{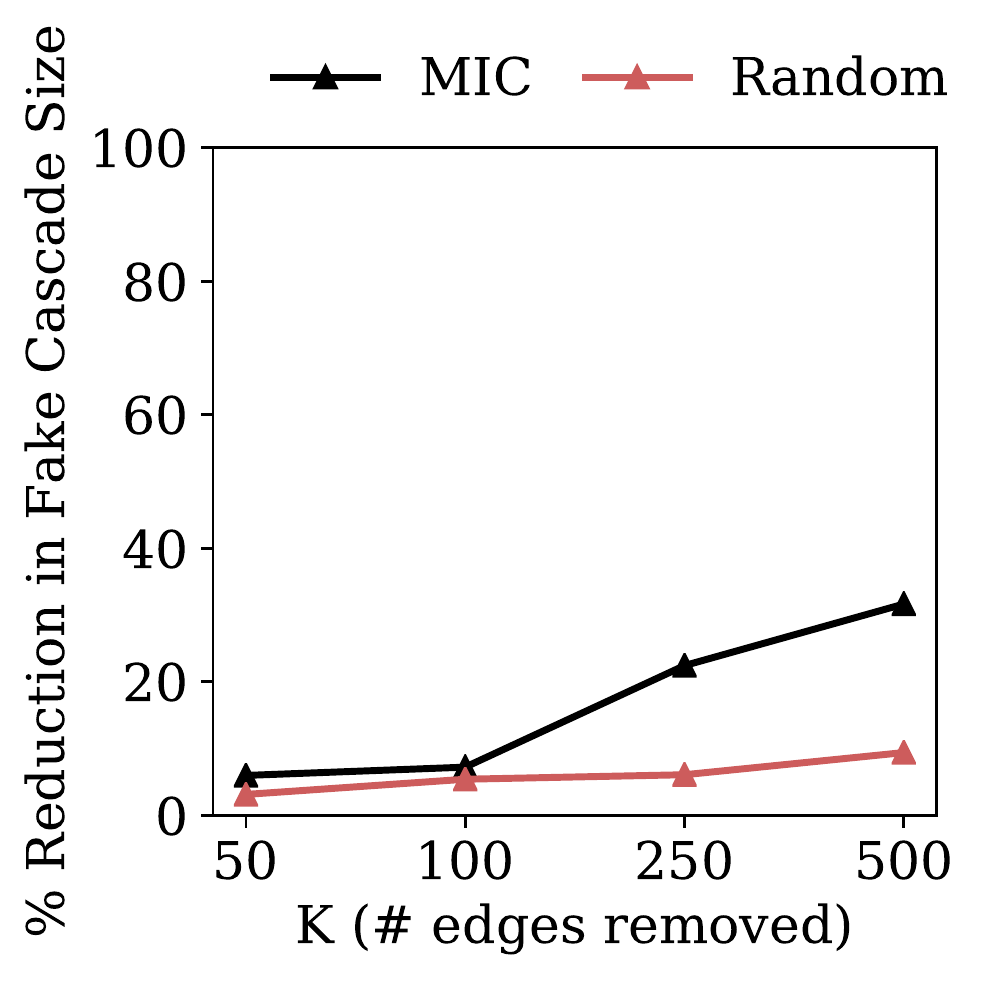}
    \caption{Intervention: Twitter-2}
    \label{fig:einter_tma}
    \end{subfigure}
    ~
\caption{Intervention Analysis on Twitter-1 and Twitter-2 (a, b) Node Interventions (c, d) Edge Interventions.}
\label{fig:interventions}
\end{figure*}

\subsection{Characteristics of Influential Users}

Table~\ref{tab:userfeats} lists the features of users identified as most influential for true and fake news under estimated parameters; reported \# of followers, posts from 2009 Twitter-1 snapshot. 

\noindent \textbf{Features of Inf(T):} Inferred influential users identified for true news, as seen, correspond largely to accounts of known credible news and opinion websites and blogs. In terms of topic distribution, the dominant types of influential users include accounts disseminating news related to politics, entertainment, infotainment, technology updates, and local news; and tend to have large number of direct followers.

\noindent \textbf{Features of Inf(F):} Top influential users identified for fake news include accounts with relatively fewer counts of direct followers, compared to those for true news users. For some of these the screen name and description is unavailable from TwitterAPI (reported as `unk.' in the table). Several of these accounts also do not have a listed description along with their screen name, unlike in the previous case of influential true news users. Therefore, we list the topic of the fake news cascades in which the users appear and their total count of engagements/appearances in the fake cascades.

The accounts influential in fake news propagation also appear among a diverse range of topics; similarly dominated by politics, technology, entertainment, and news or trending topics such as SwineFlu and current events. Interestingly, the identified top influential users appear among the larger and more viral fake cascades in the dataset such as ones corresponding to SwineFlu, Obama's citizenship status, LadyGaga's gender identity and technology rumors like launch of Xbox720. \texttt{BuzzFeed} interestingly has been historically linked to unreliable journalism, especially before 2014. It appears in connection with false stories related to man-eating catfish, BigFoot and other viral false stories.

\subsection{Intervention Analysis on Nodes and Edges}

In Fig~\ref{fig:interventions}  we investigate different intervention mechanisms (mechanisms to monitor or intercept the propagation paths of fake news) leveraging the inferred diffusion dynamics, so as to limit the spread of fake news on a network.

\noindent \textbf{Node Intervention:} In node intervention, we determine which nodes can be monitored, in order to block false contents from spreading in the network. The inferred influential users for fake news identified earlier are chosen candidates for node intervention under MIC, ranked by influence.

For offline evaluation of the intervention strategy, we utilize the available fake cascades in the datasets. First, we consider that $K$ users are selected for intervention/monitoring. 
If a fake news cascade reaches any of the monitored users, it can be intercepted and removed from the network, thereby limiting its future spread. The effectiveness of the interception can be evaluated based on the \% reduction in fake cascade size due to the intervention. In Fig~\ref{fig:ninter_kwon} and \ref{fig:ninter_tma}, we evaluate the proposed MIC intervention against the previously considered baselines; and we include an additional baseline \textbf{TopU} that intercepts users ranked by their total engagement count in the set of observed cascades. For the other baselines, the selection of $K$ users is as follows: rank users by their total engagement count in the cascades predicted as fake news cascades by the baseline method.

\noindent \textbf{Edge Intervention:} In edge intervention, we select $K$ edges in the network in order to intercept the propagation of fake cascades. The edges are ranked by the weight (strength of influence) $p_e^F$ under the inferred fake component of MIC. These are the identified high transmission paths for fake news cascades and thus removed/blocked. 

For offline evaluation, we again compare the percentage reduction in fake cascade size due to edge intervention with MIC, against a Random strategy that selects edges uniformly at random from the network for intervention, as shown in Fig~\ref{fig:einter_kwon} and \ref{fig:einter_tma}. Here the reduction is calculated over the size of fake cascades simulated over 1000 rounds under the fake component with and without the $K$ edges removed/intercepted for intervention. The simulations are triggered from seeds sampled from users at the head of the sequence of observed fake cascades in the datasets.

\section{Conclusion}
\label{sec:futurework}

In this work, we proposed a mixture of independent cascade models (MIC) to express and infer the diffusion dynamics of false and legitimate contents. With statistical analysis on real datasets, we confirmed notable differences in user behaviours towards fake and true contents in temporal and structural aspects of diffusion, that can be expressed with MIC. Based on that, we derived an unsupervised inference method for parameter estimation from observed unlabeled cascades, and conducted experiments on Twitter datasets with fake/ true news cascades. The experiments revealed interesting analysis of the characteristics of users identified as influential in true and fake content propagation, under the inferred diffusion dynamics; and their effectiveness towards node and edge interventions to limit fake news.

\subsection{Discussion and Future Work}

We assumed two sets of parameters $\theta_T, \theta_F$ to differentiate fake from true cascades, based on verifying that (i) differences in diffusion patterns of the two types are statistically significant in the datasets, and (ii) the datasets are built from collections of events reported during a specific period, with samples across types collected from the same data source; and no known collection biases across types.

In order to account for multiple types (such as satire, differences  in  political  stance,  source  credibility  or  content), the  mixture  model \emph{easily}  generalizes  to  multiple types of cascades, when $k > 2$ components are initialized in Algorithm 1 (the derivation is written for the general case $k$).


A limitation of the current work is that it assumes a \emph{fixed} number of components $k$, which need not be known a priori.  In future work, this can be addressed to adaptively split and merge components starting with a large initial k, while optimizing for likelihood of the cascades. In the experiments on influential users identified based on the inferred diffusion parameters, we find that the inferred set of parameters are correlated with the two types assumed in this work, in terms of engagements with fake and true cascades (Fig~\ref{fig:user-analysis}) and reduction in fake cascade size (Fig~\ref{fig:interventions}). However, although the proposed model directly generalizes to $k >2$, we consider evaluating the model on multiple types with $k$ unknown a priori for future work with multi-label datasets. 

The runtime analysis details of the inference algorithm are provided in the \emph{Appendix}.
The runtime scales in the order of $\mathcal{O}(k|C|V^2)$ which is reduced to $\mathcal{O}(k|C|VW)$ by setting a constant window $W$ smaller than $V$, where $W$ is the window size described in  Relaxation section under Parameter Estimation, $V$ is the number of users, $k$ is the number of components, and $C$ is the set of cascades. This is a limitation of applying the algorithm to large-scale graphs. In future work, we can integrate dimensionality reduction techniques to reduce the number of unique user representations.

There are other possible directions of future work. The first is to provide online estimation of parameters for time evolving networks; to allow for changing dynamics due to social bots and fake accounts with manufactured and evolving social connections. A second interesting direction is to leverage diffusion network inference to better understand polarization and existence of echo chambers, and its impact on the spread of misinformation - whether polarization fuels misinformation, and can interventions to mitigate one phenomenon support the other phenomenon.

\section{Appendix}
\label{appendix}


\subsection{Proof for Theorem 1.}

\begin{proof} Each edge (coordinate) $j$ has associated bernoulli variables $x_j^i$ with parameter $p_j^i$ for component $i$ in the $k$-component mixture distribution. The pairwise coordinate means then are defined as follows,
\begin{equation}
\label{eqn:corr}
corr(j,j') = \mathbb{E}[x_j x_{j'}] = \sum_{i=1}^{k} \pi^i p_j^i~p_{j'}^{i}\textrm{,\hspace{0.3cm}}~1 \leq j < j' \leq m
\end{equation}
The sample estimate of $corr(j, j')$ can be obtained directly from the observed live-edge graphs of unlabeled cascades. By the reduction to learning mixtures of discrete product distributions, given the sample estimates of the pairwise coordinate means, the parameters $p_j^i$ and $\pi$ can be estimated using algorithm Weights and Means (WAM) \cite{feldman2008learning} for learning mixture distributions. We restate lemmas in \cite{feldman2008learning,chen2016robust} used in the proof for completeness, with notations used in the reduction.
\begin{lemma}[\cite{feldman2008learning}]\label{wam}
For $k=\mathcal{O}(1)$ and any $\epsilon', \delta'>0$, WAM runs in time poly~$(m/\epsilon') \cdot \log (1/ \delta')$ and outputs a list of poly~$(m/\epsilon')$ many candidates, at least one of which (with probability at least $1-\delta'$) satisfies the following,
\begin{equation*}
|\hat{\pi}^i - \pi^i| \leq \epsilon' , \forall i
\quad\mathrm{and}\quad
|\hat{p}_e^i - p_e^i| \leq \epsilon', \forall \pi^i \geq \epsilon'
\end{equation*}
\end{lemma}

\begin{lemma}[Lemma 4 in \cite{chen2016robust}]
Given graph G and parameter space $\vartheta$ such that $\forall \theta_1, \theta_2 \in \vartheta$ , ${||\theta_1 - \theta_2||}_{\infty} \leq \epsilon_0 $, then, $\forall S \subseteq V$,
\[
|\sigma_{\theta_1} (S) - \sigma_{\theta_2} (S)| \leq mn \epsilon_0
\]
\end{lemma}
Using the above lemmas and setting $\epsilon_0 = \frac{\epsilon}{mn}$, $\delta'=\delta$ and $\epsilon' =  \frac{\epsilon}{mn}$, the sample complexity for the desired influence function estimate is obtained. WAM requires sample estimates for $\mathbb{E}[x_j x_{j'}]$ for all $1 \leq j < j' \leq m$ to be within an additive accuracy of $\epsilon_{matrix} = \left(\frac{\epsilon'^2}{m^2}\right)^{(k+1)}$. $x_j x_{j'} \in \{0,1\}$ and therefore is Bernoulli distributed with some parameter say $p_{jj'}$ equal to $\mathbb{E}[x_j x_{j'}]$. Let $\hat{p}_{jj'}$ be the sample estimate for $\mathbb{E}[x_j x_{j'}]$ calculated from the observed cascades. Since each observed cascade is independently generated, we can compute the sample complexity of estimating $\mathbb{E}[x_j x_{j'}] = p_{jj'}$ within additive accuracy of $\epsilon_{matrix}$ given the observed cascades. Applying chernoff bounds, we get $P(|\hat{p}_{jj'} - p_{jj'}| \ge \epsilon_{matrix}) \leq \delta_{matrix}$ with number of observed samples being at least $\frac{2+\epsilon_{matrix}}{\epsilon_{matrix}^2} \ln \frac{2}{\delta_{matrix}}$. Applying union bound, we get $P(|\hat{p}_{jj'} - p_{jj'}| \ge \epsilon_{matrix}) \leq \delta_{matrix}m(m-1)/2$ for all $j,j' \in [m]$. Setting $\delta_{matrix} = \frac{2\delta}{m(m-1)}$, we get with probability at least $1-\delta$, $|\hat{p}_{jj'} - p_{jj'}|$ is within additive accuracy of $\epsilon_{matrix}$ for all $j, j'$ and the sample complexity is $\mathcal{O}\left((\frac{n^4 m^8}{\epsilon^4})^{k+1} \ln \frac{m}{\delta} \right)$.

\end{proof}

\begin{figure}[t]
\centering
\begin{subfigure}[t]{0.43\columnwidth}
     \includegraphics[height=4cm,width=\columnwidth]{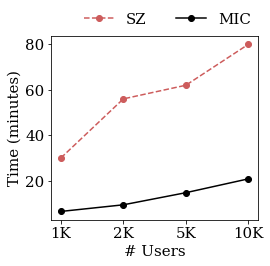} 
    \caption{Runtime analysis of minutes vs. \# of active users considered in Twitter-2.}
\end{subfigure}
\begin{subfigure}[t]{0.53\columnwidth}
     \includegraphics[height=3.7cm,width=\columnwidth]{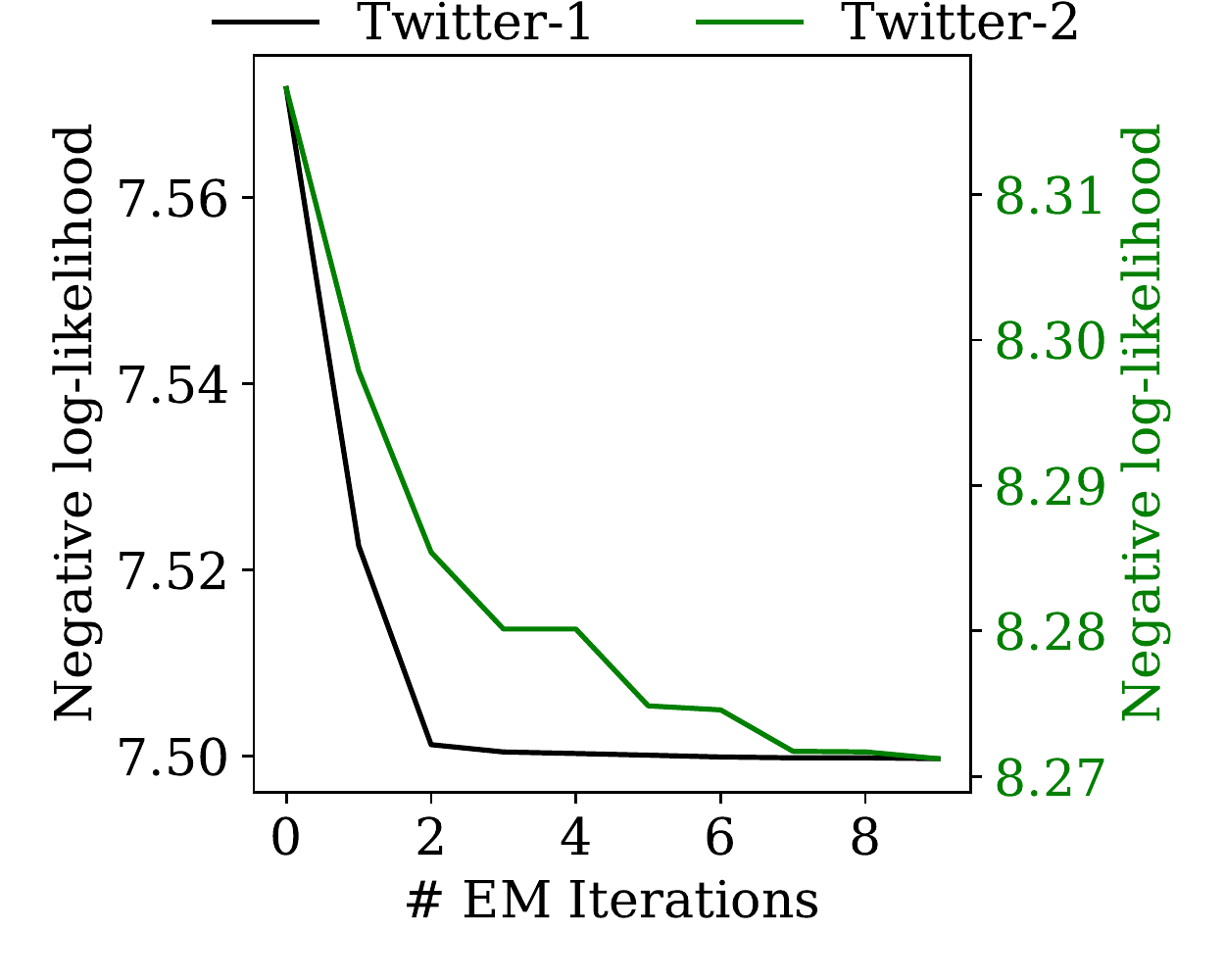} 
    \caption{Negative log- likelihood per cascade vs. \# iterations in Twitter-1 and Twitter-2.}
\end{subfigure}
\caption{Runtime Analysis.}
\label{fig:runtime}
\end{figure}

     

\subsection{Runtime Analysis}
In Fig~\ref{fig:runtime}, the runtime analysis of MIC vs. baseline SEIZ (SZ) on Twitter-2 are provided. The baseline SEIZ is run with time interval of 24hours and cut-off time of 10K hours, and it runs differential equation solvers for each cascade, to fit the data with parameters specific to each cascade. The runtimes are evaluated and compared on Intel(R) Xeon(R) CPU E5-2630 v3 @ 2.40GHz on single thread in python. Multi-threading, parallelization is left to future implementations. 

The runtime analysis is conducted on Twitter-2, since is the larger of the two datasets (with more users and more cascades), so that runtime can be analyzed with respect to different user sizes. We implemented vectorized computations and pre-computed users and cascades needed in the likelihood computation at the start of the EM iterations which improves computational efficiency, and reduces impact of number of cascades on runtime due to vectorization.


The EM estimation in Algorithm 1 is trained till convergence, i.e. the change in likelihood is smaller than $0.01$. The lookback window $W$, discussed in Section Relaxation under Parameter Estimation, is set to 10 past events. The value of $W$ impacts computational time and should be set to a constant smaller than $V$, that is the number of users. In the experiments, we set $W$ with line search in the range $\{5, 10, 15\}$ based on cross validation for computational efficiency.

The EM converges within few iterations. The worst-case runtime complexity per EM iteration is $\mathcal{O}(k|C|V^2)$ where $V$ is the number of users and $C$ is the set of cascades and $k$  is the number of components, and by setting $W$ to a constant smaller than $V$, the complexity reduces to $\mathcal{O}(k|C|VW)$.

\subsection{Additional Data Statistics}

We provide the follower graph statistics in Table~\ref{tab:follgraph} available in Twitter-1 \cite{kwon2013prominent}. The follower graph was used for structural diffusion analysis; and is a directed graph between the active users considered in the dataset, which appear at least five times in the cascades set; as described in the Section on real-world datasets. The direction of the edge from A to B indicates that A follows B. The table provides degree distribution and connected components statistics.

\begin{table}[t]
    \centering
    \begin{tabular}{l|r}
    \toprule
    Follower Graph & Value \\
    \midrule
    \# Edges & 27K \\
    \# Active users & 3K \\
    Avg Out-degree & 6.54 \\
    Max Out-degree & 126 \\
    Median Out-degree & 3 \\
    Avg In-degree & 6.55 \\
    Max In-degree & 137 \\
    Median In-degree & 2 \\
    \# strongly connected comps & 810 \\
    \# weakly connected comps & 35 \\
    \bottomrule
    \end{tabular}
    \caption{Follower Graph Statistics in Twitter-1.}
    \label{tab:follgraph}
\end{table}

\subsection{Experimental Results on Synthetic Datasets}

\begin{figure}[t]
\centering
\includegraphics[width=0.9\columnwidth]{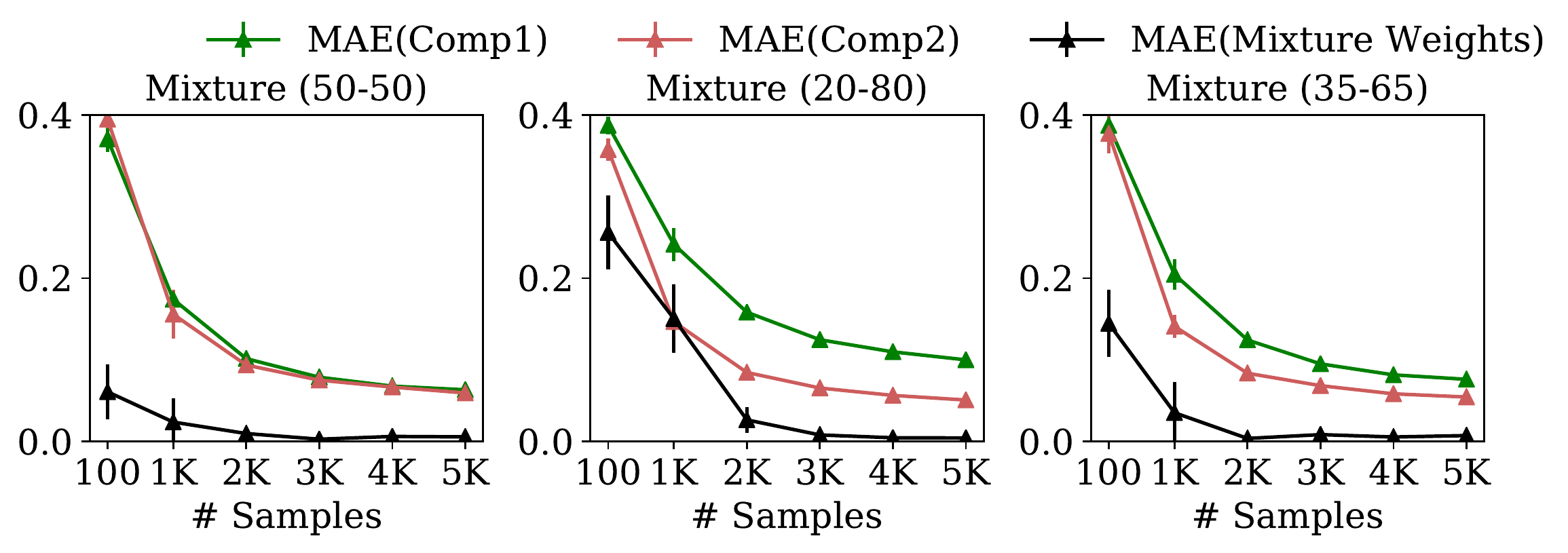}
    \caption{\emph{Parameter Recoverability.} Mean absolute error on estimated diffusion mixture model (MIC) parameters; variation with \# of cascades at different mixture distributions.}
 \label{fig:paramrec}
 ~
 \includegraphics[width=0.9\columnwidth]{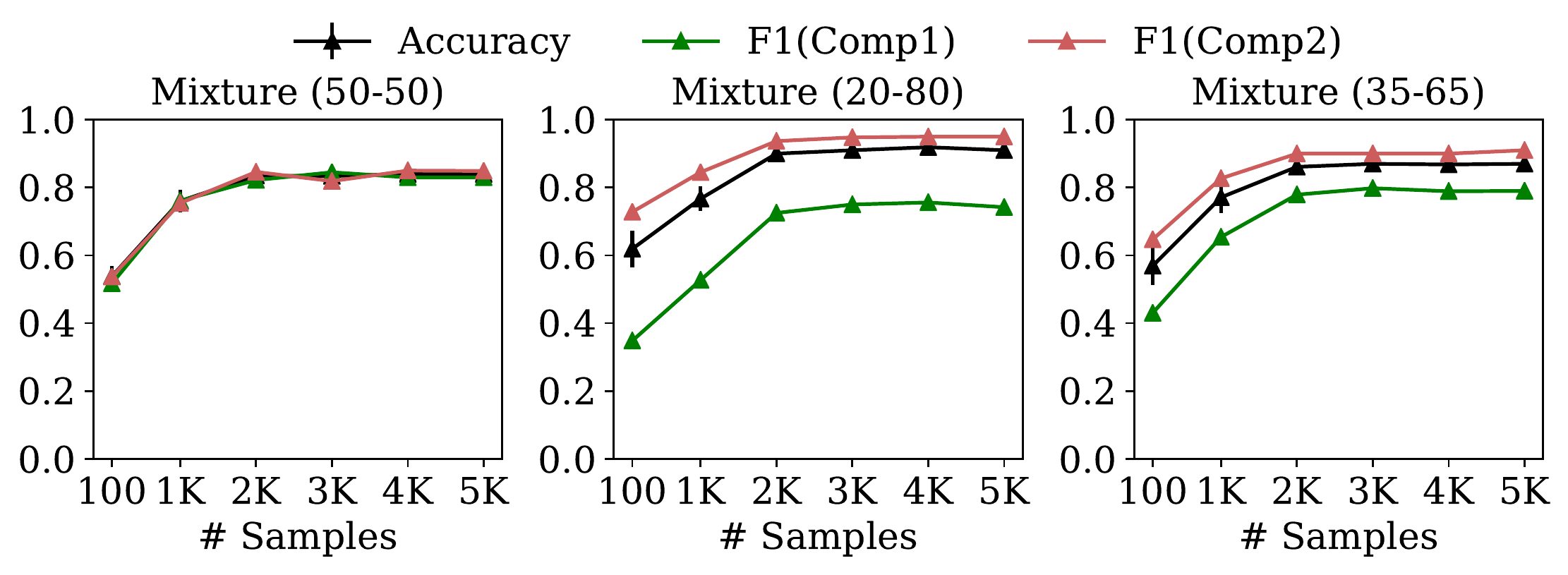}
  \caption{\emph{Cascade Separability.} Clustering accuracy and f1 results with estimated diffusion mixture model (MIC); variation with \# of cascades at different mixture distributions.}
 \label{fig:cassep}
 
\end{figure}

We construct observed cascades at different mixture distributions $\pi$ i.e. $(0.5, 0.5)$, $(0.2, 0.8)$  and $(0.35, 0.65)$. on a random graph with 512 nodes and 1024 edges and uniform [0,1] edge probabilities. Results are shown in Fig.~\ref{fig:paramrec} and \ref{fig:cassep}.

In Fig~\ref{fig:paramrec}, we evaluate parameter recoverability using mean absolute error on the
estimated diffusion mixture model (MIC) parameters on synthetic data; examining the variation with \# of cascades at different mixture distributions. In Fig~\ref{fig:cassep}, we evaluate cascade separability based on clustering accuracy and f1 with estimated diffusion mixture model (MIC); varied over \# of cascades at different mixture distributions.

\bibliographystyle{aaai}
\bibliography{filebib.bib}

\end{document}